\documentclass[conference]{IEEEtran}
\usepackage{graphicx} % Required for inserting images
\usepackage{longtable}
\usepackage{url}
\usepackage{amsmath}

\title{Are Voters Willing to Collectively Secure Elections? Unraveling a Practical Blockchain Voting System}
\author{
    \IEEEauthorblockN{Zhuolun Li\IEEEauthorrefmark{1}, Haluk Sonmezler\IEEEauthorrefmark{1}, Faiza Shirazi\IEEEauthorrefmark{1}, Febin Shaji\IEEEauthorrefmark{1}, \\ Tymoteusz Mroczkowski\IEEEauthorrefmark{1}, Dexter Lardner\IEEEauthorrefmark{1}, Matthew Alain Camus\IEEEauthorrefmark{1}, Evangelos Pournaras\IEEEauthorrefmark{1}}
    \IEEEauthorblockA{\IEEEauthorrefmark{1}School of Computer Science, University of Leeds 
    \\sczl@leeds.ac.uk, halukson@icloud.com, fshirazi710@gmail.com, febinshaji2@gmail.com, \\ tymoteuszmroczkowski@outlook.com, dex.lardner@gmail.com, mattalcamus@gmail.com, e.pournaras@leeds.ac.uk}
}

\begin{document}

\maketitle

\begin{abstract}
Ensuring ballot secrecy is critical for fair and trustworthy electronic voting systems, yet achieving strong secrecy guarantees in decentralized, large-scale elections remains challenging. This paper proposes the concept of collectively secure voting, in which voters themselves can opt in as secret holders to protect ballot secrecy. A practical blockchain-based collectively secure voting system is designed and implemented. Our design strikes a balance between strong confidentiality guarantees and real-world applicability. The proposed system combines threshold cryptography and smart contracts to ensure ballots remain confidential during voting, while all protocol steps remain transparent and verifiable. Voters can use the system without prior blockchain knowledge through an intuitive user interface that hides underlying complexity. To evaluate this approach, a user testing is conducted. Results show a high willingness to act as secret holders, reliable participation in share release, and high security confidence in the proposed system. The findings demonstrate that voters can collectively maintain secrecy and that such a practical deployment is feasible. 
\end{abstract}

\section{Introduction}
Ballot secrecy is fundamental to fair and trustworthy elections. In blockchain-based voting systems, this requirement becomes challenging due to the transparency of blockchains. Most existing blockchain voting proposals encrypt the ballots and rely on trusted authorities to hold decryption keys. However, these centralized trust models introduce vulnerabilities and concerns about misuse of privileged information. 

Prior research on perfect ballot secrecy~\cite{self-tallying-PBS} has shown that distributing secrecy responsibility to voters themselves can eliminate these trust bottlenecks. Yet, such designs typically assume that all voters will participate fully and honestly in protecting the secret, which is impractical for large-scale elections given real-world voter availability and motivation.

We propose a collectively secure blockchain-based voting system that provides flexibility for voters to decide whether to be secret holders or not. The set of secret holders, voters who opt in to help protect secrecy, each holds a partial decryption share of the encrypted ballots. Only when a sufficient threshold of secret holders release their shares at the end of the voting period can the ballots be decrypted and tallied. 

Our design also tackles a practical barrier often overlooked in blockchain voting proposals, that is, the need for ordinary voters to navigate complex blockchain interactions. Our system is designed to be as simple to use as any familiar online voting platform, requiring no prior blockchain experience. 

Beyond system design, we investigate a fundamental question: are voters willing to take on the responsibility of collectively securing elections by acting as secret holders? With a user testing, our findings show a high opt-in rate, reliable secret share release, and increased user trust in ballot secrecy and integrity, providing early evidence that voters are willing and able to actively secure elections. 

The contributions of this paper include a practical, easy to use, and scalable digital voting system design that is adaptable to different voting rules; an open-source implementation of the proposed system; a real-world deployment is used to collect data from a novel user study to understand voters behavior; and a utility function to predict the participation of voters in securing ballots, providing insights into how to incentivize contributions to security of elections.

% In the remaining paper, Section 2 provides background and related work on the status of blockchain-based voting systems. Section 3 presents the concept of collectively secure voting and proposes a blockchain-based design for it. Section 4 presents the data we collect from user testing participants to show the viability of the proposed system. Section 5 summarizes the paper and outlines potential future work.

\section{Related Work and Background}
Decentralization positively impacts trust, robustness, and ballot integrity in e-voting systems~\cite{swiss-voter-trust, bev-slr, voter-trust-factors, helbing2023democracy, POURNARAS2020160}. In decentralized voting, multiple parties may share responsibility for securing ballots~\cite{self-tallying-PBS, helios}, or tallying votes through self-tallying mechanisms in which ballots are posted to a public bulletin board~\cite{self-tallying-PBS, public-bulletin-board} for transparency. A public blockchain can serve as a public bulletin board, ensuring ballot integrity and public verifiability. As a result, a number of blockchain-based voting proposals have emerged~\cite{bev-slr}.

\subsection{Challenges in Blockchain-Enabled Voting Systems}
Publishing ballots to a public blockchain~\cite{tivi, follow_my_vote} allows monitoring of election in real time. This threatens fairness, as early partial tallies can influence remaining voters~\cite{ballot-secrecy-matters, Helios-with-ballot-privacy}, potentially changing the election outcome~\cite{proposed-crypto}. To mitigate this, cryptographic techniques such as cryptographic commitments~\cite{bev-with-commitment}, homomorphic encryption~\cite{bev_hm_enc_1, bev_hm_enc_2, bev-with-homomorphic-encrypiton-1}, zero-knowledge proofs~\cite{smart-contract-boardroom, smart-contract-boardroom-with-snark}, and secret sharing schemes~\cite{bev_secret_sharing_1, proposed-crypto} are adopted to keep ballots confidential during voting while enabling public verifiability of results~\cite{bev-slr}. However, these cryptographic designs introduce further challenges, outlined below.

\emph{Level of Ballot Secrecy:}
The strength of ballot secrecy varies depending on who controls the decryption keys. For example, some homomorphic encryption~\cite{bev_hm_enc_1, bev_hm_enc_2, bev-with-homomorphic-encrypiton-1} or secret sharing~\cite{bev_secret_sharing_1} schemes rely on a trusted authority. The authority may not share the same incentive structure as voters and can be vulnerable to coercion and corruption. Survey-based studies have shown consistently low public confidence in external election organizations, whether government bodies~\cite{ballot-secrecy-preceptions, UK-declined-government-trust} or commercial organizations~\cite{voter-trust-factors}. 

Allowing voters to protect ballots could enhance trust and increase voter participation~\cite{swiss-voter-trust, warkentin2018social}. When all voters participate, this idea is formalized in the notion of perfect ballot secrecy~\cite{self-tallying-PBS}, which guarantees that no partial tally can be learned unless all remaining voters cooperate. However, such designs~\cite{bev-with-homomorphic-encrypiton-2, self-tallying-PBS} come with practical limitations. All voters are required to follow the protocol involving multiple actions. If any voter fails to follow, the remaining voters run recovery protocols, adding further complexity for voters. As a result, these proposals are only suitable for small-scale elections.

\emph{Adaptability to Different Voting Scenarios:}
An ideal voting system should support a broad range of voting scenarios, such as multiple-choice ballots, ranked ballots, and participatory budgeting ballots. However, the cryptographic designs can constrain the kinds of ballots that the system can support. For example, the system proposed by McCorry et al.~\cite{smart-contract-boardroom, smart-contract-boardroom-with-snark} uses zero-knowledge proofs to ensure ballot validity. However, the proposals only support ballots that can be encoded as a binary value. In their proposals, a tallying party brute forces the sum of the encrypted ballot values. Similarly, systems relying on additive homomorphic encryption~\cite{bev-with-homomorphic-encrypiton-1, bev-with-homomorphic-encrypiton-2, bev_hm_enc_1, bev_hm_enc_2} can only compute numeric sums of encrypted ballots.

\emph{Public Perception and Practicality:}
Voters lack confidence in blockchain voting systems and may not fully understand how their privacy is preserved~\cite{ballot-secrecy-preceptions, BEV-preception}. Moreover, most existing proposals remain theoretical~\cite{bev-slr} and have little attention to practicality and user experience. Additionally, most public blockchains remain difficult for the general public to use~\cite{blockchain-user-experience}. Tasks such as wallet creation, secret key management, and funding transactions can be challenging for voters.

\emph{Comparison of Blockchain Voting System:}
Table~\ref{tab: comparison} compares different categories of blockchain-based voting solutions. For the proposed system, ballot secrecy is rated as medium to high, considering that secret holders are a subset of voters, and the adopted cryptographic protocol provides verifiability to increase the cost of breaching ballot secrecy for secret holders. 

\begin{table*}[]
\centering
\caption{Comparison of blockchain-based voting solutions categorized by the use of cryptographic approaches for ballot secrecy}
\begin{tabular}{p{0.32\linewidth}p{0.11\linewidth}p{0.2\linewidth}p{0.15\linewidth}p{0.09\linewidth}}
\hline
Approach                                                                                                                        & Ballot Secrecy & Supported Voting Rules & Simplicity for Voters & Scalability                            \\ \hline
HM ENC (external key holder)~\cite{bev-with-homomorphic-encrypiton-1,bev_hm_enc_1,bev_hm_enc_2} & Low            & Adding numeric ballots                 & High                  & Medium   \\
HM ENC (voters as key holders)~\cite{bev-with-homomorphic-encrypiton-2}                                           & High           & Adding numeric ballots                 & Low                   & Low \\
Secret sharing (external key holder)~\cite{bev_secret_sharing_1}                                          & Low            & Any ballots any tallying rules                 & High                  & High                                   \\
Secret sharing (voters as key holders)~\cite{self-tallying-PBS}                                                          & High           & Any ballots any tallying rules                   & Low                   & Low                                    \\ 
Zero-knowledge proof~\cite{smart-contract-boardroom, smart-contract-boardroom-with-snark}                                  & High           & Adding binary ballots                & High                  & Low                    \\ \hline
Proposed System                                                                                                               & Medium - High  & Any ballots any tallying rules                   & High                  & High                                \\  \hline

\end{tabular}
{\raggedright 
\begin{itemize}
    \item HM ENC: Homomorphic encryption \\
    \item Secret sharing approaches support any voting rules because each individual ballot is decrypted when tallying.
\end{itemize}
\par}
\label{tab: comparison}
\end{table*}

\subsection{Background: Blockchain-Based Timed-Release Encryption}
We adopt the smart contract based timed-release encryption method proposed by Li et al.~\cite{proposed-crypto} to encrypt ballots. Ballot messages are protected until the designated decryption time. Given $n$ secret holders and a threshold $t$, the protocol allows a secret $k$ to be shared among the $n$ secret holders, such that any $t$ out of $n$ secret holders to recover the secret $k$. The reveal-verifiability property of the secret sharing protocol ensures that when $k$ is disclosed, the correctness of the secret shares can be publicly verified and that any misbehavior is detectable.

The protocol consists of the following phases: (i) \emph{Setup}: Each secret holder generates a key pair and publishes the public key to the smart contract. (ii) \emph{Message encryption}: A client encrypts a message given the public key of the secret holders. Along with the ciphertext, the client posts necessary cryptographic information $P$~\footnote{$P$ is used to compute and verify secret shares. The cryptographic details are omitted here for succinctness. $P$ is corresponded to $(g_1^r, g_2^r, \alpha_t, ..., \alpha_n)$ in the original paper.} to the smart contract. Each secret holder uses $P$ to compute its secret share and keeps it confidential. (iii) Secret recovery: At designated time, secret holders publish secret shares. With a threshold of valid shares revealed, anyone can reconstruct the original message and verify that it was decrypted honestly using $P$.

% \begin{itemize}
%     \item Support for incentives and penalties: The smart contract can handle cryptocurrency deposits and provide financial incentives for secret holders who behave honestly. Rewards can be distributed automatically upon successful decryption, while penalties are enforced for misbehavior. 
%     \item Unified time reference: All protocol participants rely on the blockchain native time reference to determine when the decryption phase should begin. This prevents disputes about whether a secret was revealed too early or too late.
%     \item Automated dispute resolution: By combining verifiable timestamps and on-chain commitments, the smart contract can detect and penalize premature disclosure, off-chain collusion, or the publication of invalid secret shares. For example, if a secret holder leaks their share before the specified time or submits an incorrect share, other participants can submit cryptographic proof of misbehavior, triggering automatic penalties.
% \end{itemize}

\section{Collectively Secure Voting}
We propose the concept of collectively secure voting, which decentralizes the responsibility for ballot secrecy to the voters themselves. Unlike previous designs~\cite{self-tallying-PBS, bev-with-homomorphic-encrypiton-2} that allow participation from all voters, voters optionally opt in as secret holders. This flexibility balances strong secrecy guarantees with the practical realities of varying voter availability and willingness to participate in securing ballots.

% The smart-contract-based timed-release encryption scheme proposed by Li et al.~\cite{proposed-crypto} is used to ensure that ballots remain confidential during the voting period while remaining verifiable and tamper-resistant.

\begin{figure}[h!]
\centering
\includegraphics[width=9cm]{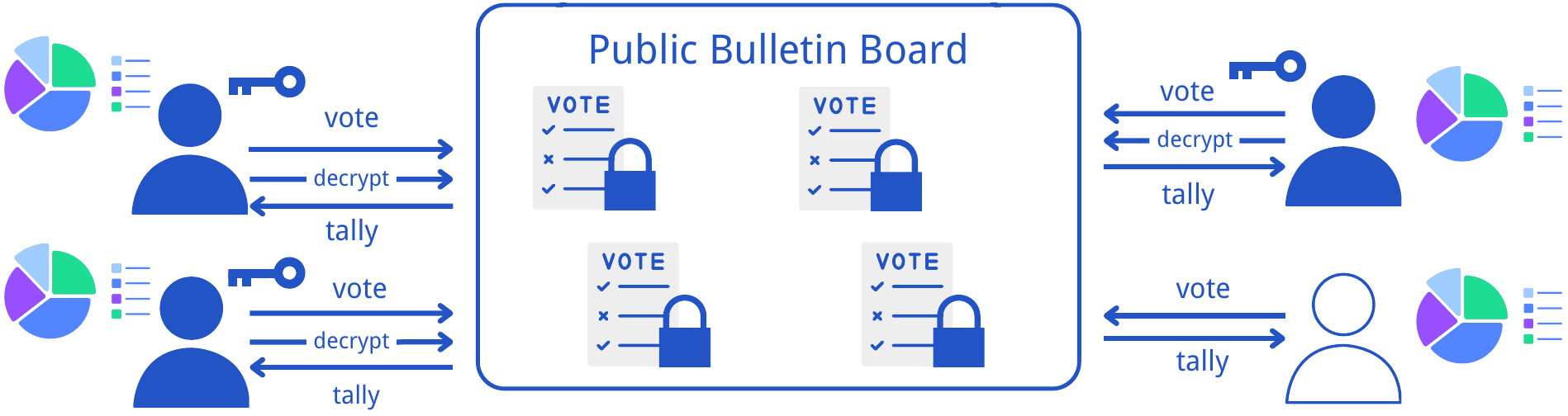}
\caption{Overview of collectively secure voting}
\label{fig:diagram}
\end{figure}

Figure~\ref{fig:diagram} provides an overview of the system architecture. All protocol interactions occur through a smart contract on a public blockchain\footnote{As the cost of employing a decentralized robust public bulletin board, the host is expected to cover the required gas fees for voters to send transactions. As of June 2025, gas costs in blockchains such as Sui remain low (approximately 0.006 USD per transaction~\cite{sui_gas_fee}). Each voter only sends one transaction and each secret holder sends one additional transaction.}. In this example, three out of four voters choose to be secret holders. All ballots are encrypted and the decryption key is shared with the secret holders. Ballots can be decrypted with any two out of the three secret shares.

\subsection{System Design}
Below is a voting life cycle of the proposed system.

\emph{Eligibility Check:}
Eligibility verification is decoupled from the voting process itself. The host checks voter eligibility off-chain and provides each eligible voter with a unique secret identifier $I$, which should be included in the ballot to show eligibility. The hash of each identifier, $h(I)$, is then hashed again and published on-chain as $h(h(I))$. $h(h(I))$ from all the voters form the set of $S$ eligible participant identifiers. This double-hashing scheme is designed to preserve the secrecy of $I$ during registration. It works as follows: (i) During registration as a secret holder, a voter submits $h(I)$. The smart contract verifies their eligibility by checking that $h(h(I))$ is in $S$. $I$ is not revealed at this stage. This is to prevent malicious parties from using $I$ to forge a ballot on behalf of the voter. (ii) During ballot casting, a voter includes $I$ in the ballot. Since the ballot remains encrypted until the tally phase, $I$ stays confidential during voting. Once decrypted, $I$ is used to verify that the ballot comes from an eligible voter.

\emph{Smart Contract Setup:}
With the set $S$ of eligible voters, the host sets up a smart contract for the voting session. The host configures the contract with parameters including the registration period, voting start and end times, allowed ballot format, and, optionally, deposit and reward for secret holders. The host also stores the eligibility set $S$ in the smart contract.

\emph{Secret Holder Registration:}
Any eligible voter can choose to become a secret holder to help secure ballot secrecy. To register at the smart contract, the voter provides $h(I)$ and a public key $pk$ generated from a one-time key pair $(sk, pk)$ created specifically for this voting session. The contract verifies that $h(h(I))$ is in $S$ before accepting the registration to ensure that only eligible voters can become secret holders. 

\emph{Ballot Casting:}
Each voter can cast their vote by encrypting their ballot using the public keys of the registered secret holders. The ballot submission includes the encrypted ballot content along with the secret identifier $I$, and the cryptographic information $P$ derived from the reveal-verifiable timed-release encryption protocol~\cite{proposed-crypto}. To resist voter coercion, the system allows each voter to overwrite their ballot by submitting a new one. Only the latest valid ballot per identifier $I$ is counted when tallying the final result.

\emph{Releasing Secret Keys:}
To tally, secret holders reveal their secret keys $sk$. The original timed encryption scheme~\cite{proposed-crypto} proposed to release the secret share of each message instead of the secret key. That is because in a general setting, the secret key is a long-term key for the secret holder serving messages at different decryption times. In the proposed system, each key pair is generated exclusively for a single election and all ballots should be decrypted at the same time, hence secret holders can simply reveal the secret keys. Secret shares of all ballots can be publicly derived given the revealed secret keys. 

\emph{Tallying and Verification:}
Once enough valid secret keys have been published to meet the decryption threshold, anyone can reconstruct the decryption key and decrypt the ballots. Valid ballots are the ones that match the required format, cast within the valid voting window, and have a secret identifier $I$ that matches the published eligibility list.

\subsection{Voter-Friendly System Implementation}
Blockchain-based applications are often associated with steep entry barriers for users. In the context of voting, where usability and accessibility are essential, such complexity is a major concern. We propose a user-friendly design that allows voters to use the blockchain-based system as easily as they would in a traditional e-voting system, while still benefiting from the security features offered by the blockchain. The implementation of our system is open-sourced~\cite{implementation}.

\emph{Zero Blockchain Interaction:}
Voters are not required to use wallets or control blockchain addresses, as they can be identified by $S$. Messages are sent to the website server to deliver to the blockchain as a proxy sender. Since all messages received by the website server are cryptographically secured, no honesty assumption is required for the website. Voters can verify that their transactions have been successfully recorded. If users observe that messages are sent to the website but not to the blockchain, they can either simply look for another proxy sender or send the messages themselves.

\emph{Client-Side Cryptography:}
Client-side cryptography is implemented to handle all blockchain-related complexities behind the scenes. Upon registering as a secret holder, a new cryptographic key pair is generated directly in browser. The secret key is automatically encrypted and stored in local storage. Only the public key is sent to the website server, which then creates and submits a blockchain transaction. Similarly, casting a ballot and submitting a secret key from the local storage are also implemented as one-click operations (see Appendix~\ref{apdx: website}).

\section{User Testing}
A user testing is conducted focusing on understanding the willingness to register as secret holders, and do secret holders reliably show up to release their secret keys. Forty participants took part in the user testing, having different education levels ranging from A-Level to Master's degrees, different professional backgrounds including but not limited to engineering, business, health and medical sciences.

% \subsection{User Testing Day}
% To best approximate real-world usage, we designed the testing session to span a full day, replicating the complete life cycle of a typical election using our system. As the host, we managed the voting session end-to-end. 

% \begin{enumerate}
%     \item Registration \& Pre-Testing Survey (00:00 - 12:00): Participants received an email with a registration link. Each participant could choose whether to opt in as a secret holder. Participants also completed a pre-test survey to assess their initial understanding and perception of the system.
%     \item Voting (12:00 – 17:00) Participants cast their votes through the system website.
%     \item Secret Submission (17:00 – 17:30) Secret holders were required to submit their secret keys to enable decryption of the ballots. To motivate reliable participation, secret holders who succeed in submitting their keys earn additional raffle tokens; secret holders who fail to do so forfeit all of their tokens.
%     \item Voting Result Reveal \& Post-Testing Survey (after 17:30) Once the threshold of shares was reached, the final result was decrypted and published. Participants could verify the outcome on the website and complete a post-test survey to reassess their understanding and perceptions of the system.
% \end{enumerate}

\begin{figure}[h!]
\centering
\includegraphics[width=9cm]{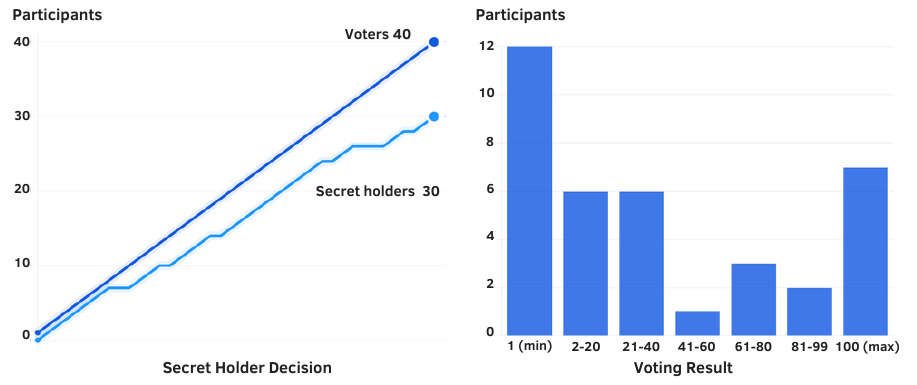}
\caption{Decision of secret holder registration in user testing (left) and voting result on the number of extra raffle tokens for successful secret holders (right)}
\label{fig: user_testing_result}
\end{figure}

\emph{High Secret Holder Registration Rate:}
Of the 40 participants, 30 chose to be secret holders. Early voters were slightly more likely to opt in as secret holders (see Figure~\ref{fig: user_testing_result} left). Although the observation is insignificant without a larger sample, it provides a potential research direction of understanding the herding effect of being secret holders.

\emph{Beyond Expected Experience:}
In both the pre-testing and post-testing surveys, participants rated how strongly they agreed that the system provides key guarantees, such as ballot integrity, voter anonymity, and simplicity, on a scale of 0 (strongly disagree) to 1 (strongly agree) that the system provides a feature. After using the system, participants have shown increased confidence that the system provides ballot integrity (from 0.65 to 0.75, p=0.037), anonymity (from 0.57 to 0.72, p=0.008), and simplicity for voters (from 0.73 to 0.81, p=0.051). This also positively suggests that blockchain-based voting systems are more likely to be accepted by the public if they are given an experience of using one.

% \begin{figure}[h!]
% \centering
% \includegraphics[width=9cm]{images/experience.png}
% \caption{A comparison of participants perception of the system before and after the user testing}
% \label{fig: user_experience}
% \end{figure}

\emph{Value of the Secret Holder Role:}
To test whether participants appreciate the secret holder role, a special question is crafted for the participants to vote on: how much extra raffle weight should successful secret holders receive compared to regular voters? In the user testing, each voter was given 10 raffle tokens to draw an Amazon voucher as a reward; they voted on how many additional tokens secret holders should get. The voting result (Figure~\ref{fig: user_testing_result} right) shows that, on average, participants allocated an extra 38.77 raffle tokens per successful secret holder, reflecting genuine subjective opinions on the value of this role. The result suggests that participants recognize the effort and responsibility involved in maintaining ballot secrecy.

\begin{figure}[h!]
\centering
\includegraphics[width=9cm]{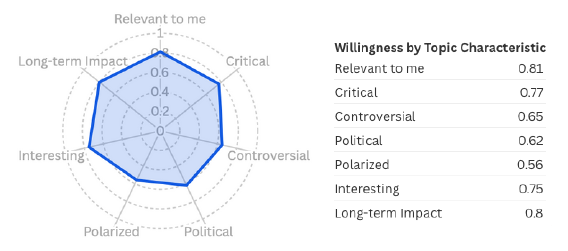}
\caption{Participants’ willingness to act as secret holders for voting topics with different characteristics on the scale of 0 (very unlikely) to 1 (very likely)}
\label{fig: willingness_by_topics}
\end{figure}

\emph{Willingness to be Secret Holder by Voting Topics:}
When asked how likely they would be to opt into voting topics with different characteristics, as shown in Figure~\ref{fig: willingness_by_topics}, participants reported greater willingness to contribute to projects with high relevancy and long-term impact; less likely to opt into controversial, polarized, and political topics. Note that the host can flexibly adjust rewards or deposits to encourage more voters to become secret holders when needed.

\emph{Factor Importance of the Secret Holder Opt-In Decision:}
Before the user study, we model the utility of acting as a secret holder as a function with factors that could positively or negatively affect a voter’s willingness to participate. The factors include \emph{monetary reward} for successful secret holders; \emph{goodwill} as the intrinsic sense of civic duty or contribution to the collective benefit of protecting ballot secrecy; \emph{obligation} as the perceived burden of the requirement to submit the secret key; and \emph{deposit} required to register as a secret holder, as the perceived cost or risk of losing for non-compliance.

% \begin{itemize}
%     \item Monetary reward: the expected benefit of receiving rewards for successfully releasing secret shares. 
%     \item Goodwill: the intrinsic sense of civic duty or contribution to the collective benefit of protecting ballot secrecy.
%     \item Obligation: the perceived burden of the requirement to submit the secret key at the correct time.
%     \item Deposit: the perceived cost or risk of losing a deposit for non-compliance or early disclosure. In the user testing, secret holders risk their raffle tokens.
% \end{itemize}

\begin{figure}[h!]
\centering
\includegraphics[width=9cm]{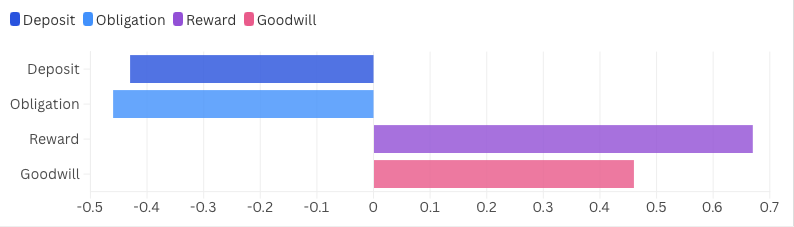}
\caption{Factor importance of the secret holder opt-in decision}
\label{fig: factor_importance}
\end{figure}

When asked to evaluate the importance weight of these factors, as shown in Figure~\ref{fig: factor_importance}, participants viewed explicit rewards as the strongest factor encouraging them to become secret holders. In contrast, deposits and strict obligations were generally perceived as discouraging factors.

To further examine how these self-reported factors relate to participants’ real decisions, a logistic regression model (Accuracy: 80.65\%) is fitted using the reported importance as predictors of the secret holder opt-in decision. Aligned with the self-reported results, the regression indicates that goodwill (coefficient: 0.88) and monetary rewards (coefficient: 0.35) have positive impacts, while obligation (coefficient: -0.1174) and penalty (coefficient: -0.3463) have negative impacts. Notably, the coefficient for goodwill is higher than that for monetary reward, suggesting that participants who rated goodwill highly were indeed more likely to be secret holders. In other words, voters’ sense of civic duty encourages them to participate in maintaining ballot secrecy. 

% \begin{table}[h!]
% \centering
% \caption{Logistic regression coefficients for secret holder decision}
% \begin{tabular}{l l l l l}
% \hline
% \textbf{Factor} & Monetary Reward & Goodwill & Obligation & Penalty \\
% \hline
% \textbf{Coefficient} & 0.3512 & 0.8789 & -0.1174 & -0.3463 \\
% \hline
% \end{tabular}
% \label{tab:logistic_regression}
% \end{table}

\section{Conclusion and Future Work}
This paper introduces a practical blockchain-based voting protocol that decentralizes the responsibility for ballot secrecy to voters. The user testing result supports that when given the option, voters are willing to collectively secure elections.

Several open challenges remain for future work. First, larger-scale real-world testing is needed to evaluate the system under diverse voting conditions. Second, incentive structures for secret holders could be further optimized and integrated with decentralized identity frameworks to enhance fairness and reduce reliance on centralized hosts for eligibility checks. By addressing these challenges, we hope to advance the development of secure and user-friendly blockchain voting systems.

\section{Acknowledgment}
This project is funded by a UKRI Future Leaders Fellowship (MR-/W009560-/1): ‘Digitally Assisted Collective Governance of Smart City Commons–ARTIO’.

\bibliography{references}
\bibliographystyle{ieeetr}

\appendices

\section{Voting System Web Interface}
\label{apdx: website}

\begin{figure}[h!]
\centering
\includegraphics[width=9cm]{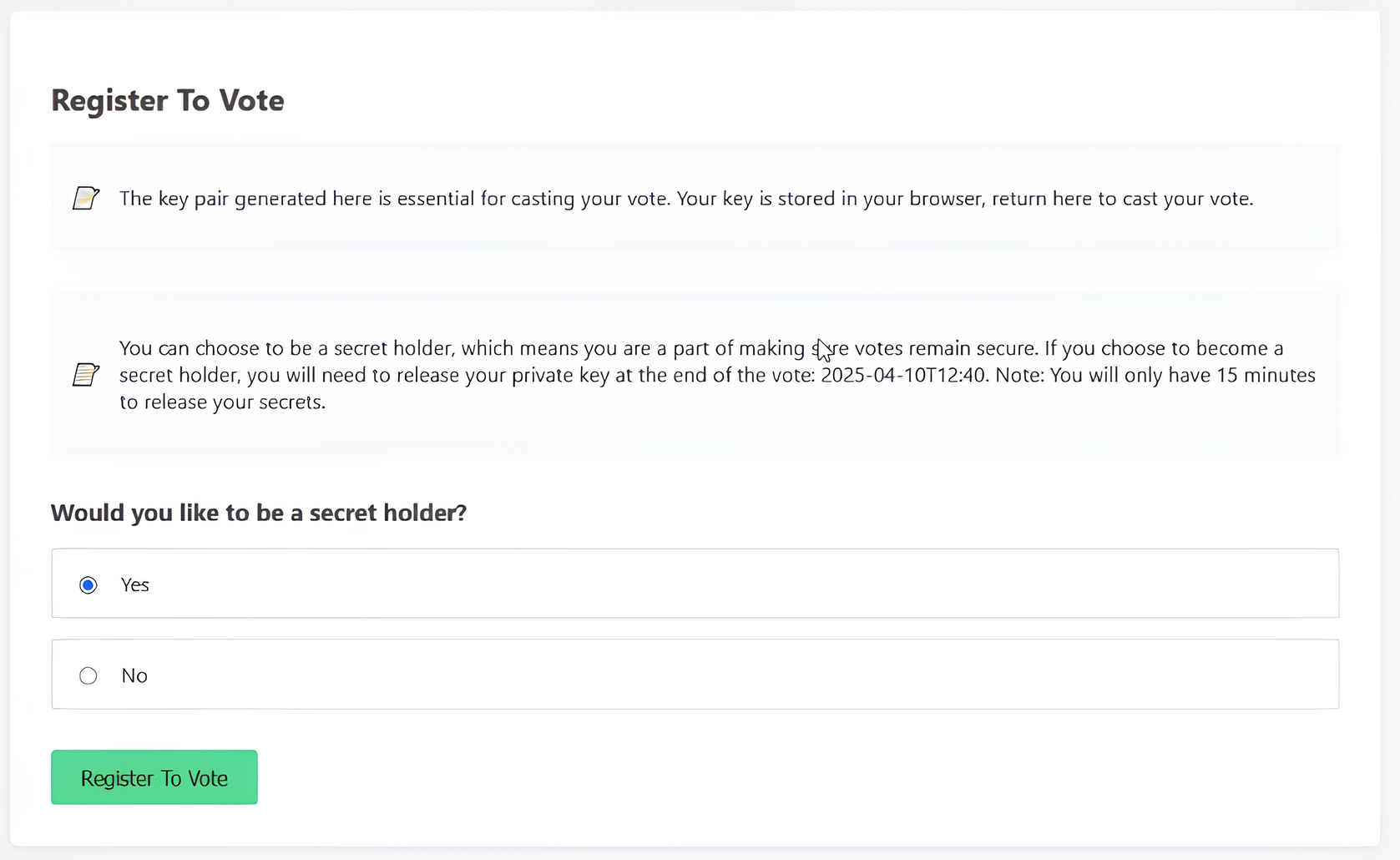}
\caption{Website user interface for registration}
\label{fig: register_screenshot}
\end{figure}

As shown in Figure \ref{fig: register_screenshot}, from the perspective of a voter, the complexity of the cryptographic operation is handled by the browser locally, leaving voters a simple interface of one-click secret holder registration. 

\begin{figure}[h!]
\centering
\includegraphics[width=9cm]{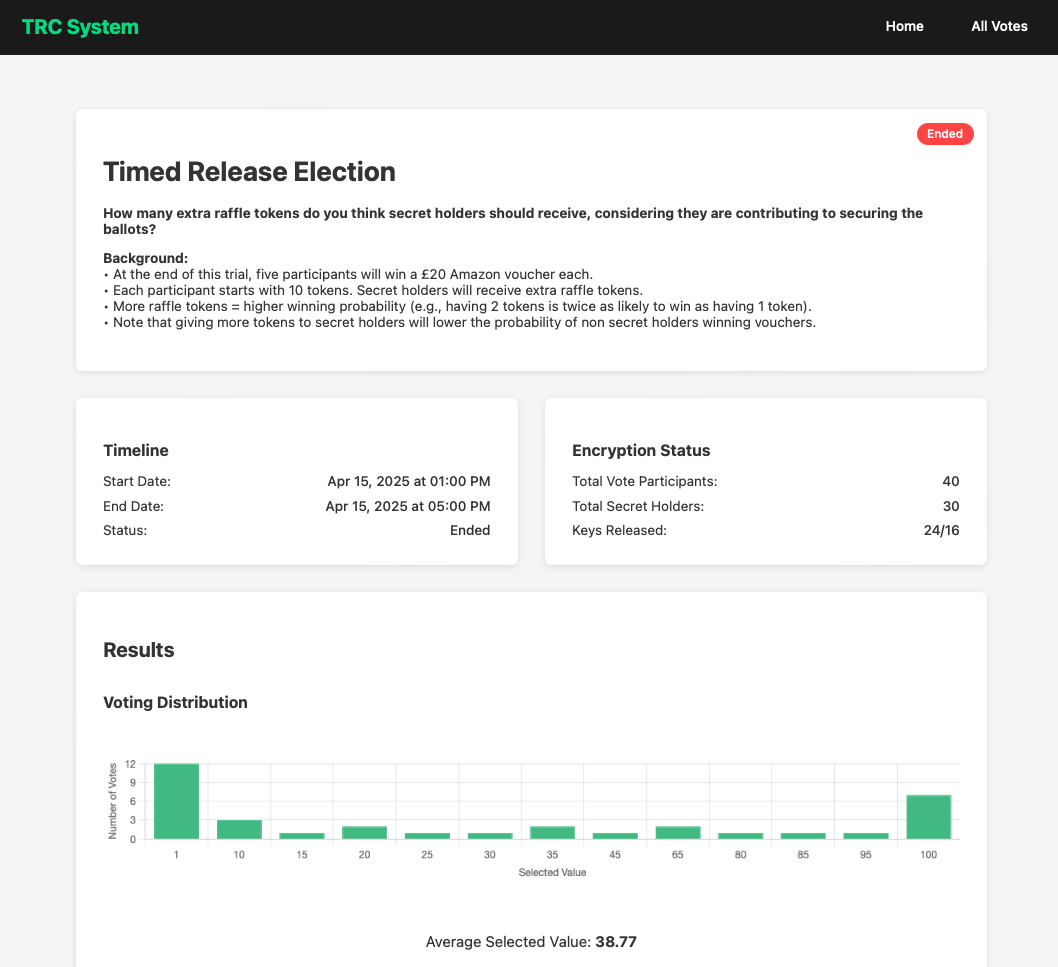}
\caption{The user testing voting webpage}
\label{fig: user_testing_screenshot}
\end{figure}

Figure~\ref{fig: user_testing_screenshot} shows a snapshot of the voting webpage displaying key voting session information and the final result. Along with the final result, a link to the blockchain explorer showing the smart contract transactions is attached for interested users to navigate and verify the result. 

% \section{User Testing Data}
% \label{apdx: user_testing_data}

% \begin{figure}[h!]
% \centering
% \includegraphics[width=9cm]{images/user_testing_result.pdf}
% \caption{Decision of secret holder registration in user testing (left) and voting result on the number of extra raffle tokens for successful secret holders (right)}
% \label{fig: user_testing_result}
% \end{figure}

\end{document}